\documentclass[%
 reprint,
 amsmath,amssymb,
 aps,
]{revtex4-2}

\usepackage{graphicx}% Include figure files
\usepackage{dcolumn}% Align table columns on decimal point
\usepackage{bm}% bold math
\usepackage{color}
\usepackage{tikz-cd}
\usepackage{amsthm}
\usepackage{amsmath}
\usepackage{mathtools}
\usepackage{enumerate}
\usepackage{amsthm}
\usepackage{xcolor}
\usepackage{pgfplots}
\usepackage[normalem]{ulem}

\definecolor{brgreen}{RGB}{0,66,7}

\usepgfplotslibrary{colormaps}
\pgfplotsset{compat=1.18}
\pgfplotsset{
/pgfplots/colormap={mymap}{rgb255(0cm)=(240,240,240); rgb255(1cm)=(0,66,7); rgb255(2cm)=(0,66,7)}}
\pgfplotsset{
/pgfplots/colormap={mymap2}{rgb255(0cm)=(240,240,240); rgb255(0.15cm)=(240,240,240); rgb255(2cm)=(0,128,255)}}

\newcommand{\dd}{\mathrm{d}}

\newcommand{\gJ}{g_{\mathcal{J}}}
\newcommand{\gF}{g_{\mathcal{F}}}
\newcommand{\J}{\mathcal{J}}
\newcommand{\F}{\mathcal{F}}
\newcommand{\B}{\mathcal{B}}

\newcommand{\X}{X}
\newcommand{\T}{\mathcal{T}}
\newcommand{\RN}{\mathbb{R}^n_{>0}}
\newcommand{\RR}{\mathbb{R}^m}
\newcommand{\img}{\mathrm{Im}}
\newcommand{\kerr}{\mathrm{Ker}}

\newcommand{\com}[1]{\textcolor{red}{#1}}
\newcommand{\eqnref}[1]{Eq. (\ref{#1})}

\begin{document}

\preprint{APS/123-QED}

\title{The Geometry of Thermodynamic Uncertainty Relations \\ in Chemical Reaction Networks}
%\thanks{A footnote to the latter title}%

\author{Dimitri Loutchko}
%\email{dimitri@sat.t.u-tokyo.ac.jp}
\author{Yuki Sughiyama}
\author{Tetsuya J. Kobayashi}
\homepage{http://research.crmind.net}
\affiliation{Institute of Industrial Science, The University of Tokyo, 4-6-1, Komaba, Meguro-ku, Tokyo 153-8505 Japan}

\date{\today}% It is always \today, today,
             % but any date may be explicitly specified

\begin{abstract}

Recently, Hessian geometry - an extension of information geometry - has emerged as a framework to naturally connect the geometries appearing in the theory of chemical reaction networks (CRN) to their inherent thermodynamic and kinetic properties.
This framework is used in this letter to derive multivariate thermodynamic uncertainty relations (TUR) for CRN.
The matrices featured in the TUR are shown to be representations of Riemmanian metric tensors, whereby one tensor characterizes the pseudo entropy production rate and the other the current fluctuations. 
It is shown that the latter tensor is a restriction of the former one to a linear subspace in the flux tangent space. 
Therefore, in addition to clarifying the geometric origin of TUR in CRN, the Hessian geometric setup yields a characterization of the error term in the TUR as the norm of a linear subspace component of the flux vector and thus characterizes the fluxes where TUR become equalities.
\end{abstract}

\maketitle

\paragraph*{Introduction.} Thermodynamic uncertainty relations (TUR) have been established as universal principles in nonequilibrium thermodynamics over the past decade.
They are as important for the understanding of nonequilibrium behaviour as the laws of thermodynamics are for the understanding of the equilibrium properties of matter.
TUR state that the inverse of the relative current fluctuations is bound by the dissipation.
In other words, they impose an energetic constraint of the precision of currents.
Thus, without the need to formulate any specific kinetic model, TUR impose energetic restrictions on biological regulatory systems such as circadian clocks \cite{gonze2006,hatakeyama2012} and proofreading kinetics \cite{banerjee2017,chiuchiu2023}.

For this reason, since their first formulation in \cite{barato2015}, TUR have been extensively explored, refined, and generalized by various methods \cite{horowitz2020}.
The first proof used large deviations theory \cite{gingrich2016,horowitz2017} but soon thereafter the information theoretical nature of TUR has been discovered and they have been reformulated as a Cramer-Rao bound \cite{hasegawa2019}, generalized to the multivariate case \cite{dechant2018} , and unified with thermodynamic speed limits \cite{van2022,van2023}.
TUR also have been formulated for time-dependent driving \cite{koyuk2020}, quantum systems \cite{hasegawa2021}, for oscillatory dynamics \cite{marsland2019}, and as time-information uncertainty relations \cite{nicholson2020time}.
By now, there are various applications such as the inference of entropy production from short trajectories \cite{manikandan2020,otsubo2020} or the determination of the Fano factor for enzymes \cite{barato2015fano}.

Chemical reaction networks (CRN) are essential for the understanding of complex biochemical phenomena \cite{mikhailov2017} but due to their nonlinear nature, their thermodynamic and kinetic properties are notoriously difficult to access \cite{rao2016,esposito2020}.
Recently, information geometric techniques have been adapted to provide a unified geometric framework which encompasses the thermodynamic, kinetic, and algebraic aspects of CRN \cite{sughiyama2021,kobayashi2021,yoshimura2021information, loutchko2022,kolchinsky2022,dechant2022,kobayashi2022graph,yoshimura2023}.
%This geometry has been termed Hessian geometry and it has been worked out for the concentration and potential spaces \cite{sughiyama2021,kobayashi2021} as well as for the spaces of fluxes and chemical forces \cite{kobayashi2022geometry,dechant2022,kobayashi2022graph}.
This geometry has been termed Hessian geometry and it provides a natural framework to tackle TUR for CRN.

Thus far, Hessian geometry has been worked out only for deterministic CRN, whereas TUR in general require a stochastic description in order to account for the fluctuations.
However, the Kramers–Moyal expansion of the chemical master equations yields a Fokker-Planck equation wherein the diffusion coefficients are given purely by macroscopic quantities.
This setup is used in \cite{yan2019,chetrite2019,yoshimura2021TUR,leighton2023} and also forms the basis for this letter.
We derive a multivariate TUR for CRN as a matrix inequality.
The matrices are representations of Riemannian metric tensors whereby one tensor characterizes the pseudo entropy production rate and the other
the inverse relative fluctuations of the chemical currents.
The inequality is a consequence of the letter tensor being a restriction of the former one to a linear subspace in the flux tangent space.
This geometrical approach to TUR has several advantages:
First, the effect of the non-diagonal terms of the diffusion matrix on the fluctuations is taken into account - this could not be achieved in previous work \cite{yan2019,chetrite2019,yoshimura2021TUR,leighton2023}.
This allows to analyze the effect of individual reactions on each of the chemical currents.
Second, the geometrical understanding of TUR offers clear conditions for minimizing fluctuations by tuning the positions of individual flux vectors.
Third, from geometry, together with the Legendre duality between chemical forces and flux vectors, an interpretation of the difference between dissipation and the inverse of relative current fluctuations in TUR can be provided as follows:
The dynamics of a CRN leaves the cycle affinities which result from external coupling to the reservoirs constant, i.e., only the orthogonal part to the linear cycle space on the force space can undergo fluctuations and this is transferred to the flux space, where fluctuations only can take place on a certain (nonlinear) subspace.
Finally, the geometric interpretation and the simplicity of the proof given here provide a blueprint on how to prove TUR beyond the Fokker-Planck approximation.
This, however, requires a setup of Hessian geometry on infinite dimensional spaces.

\paragraph*{Setup.}

We consider a CRN with $n$ chemicals $X_1,\dotsc,X_n$ and $m$ reactions $R_1,\dotsc,R_m$.
The $j$th reaction is given by
\begin{align}
    R_j: \sum_{i=1}^n S^+_{ij} X_i \rightarrow \sum_{i=1}^n S^-_{ij} X_i
\end{align}
with nonnegative integer coefficients $S^+_{ij}$ and $S^-_{ij}$.
The structure of the network is thus encoded in the $n \times m$ stoichiometric matrix $S = [S_{ij}]$ with matrix elements
\begin{align}
    S_{ij} = S^-_{ij} - S^+_{ij}.
\end{align}
The state of the reaction network is characterized by a vector of nonnegative concentration values $x = (x_1,\dotsc,x_n) \in \mathbb{R}^n_{>0}$, where $x_i$ represents the concentration of the chemical $X_i$.
The concentration space is denoted by $X := \RN$.
The deterministic dynamics of the CRN is governed by the continuity equation
\begin{align} \label{eq:dynamics}
    \frac{\dd x}{\dd t} = Sj,
\end{align}
where $j=(j_1,\dotsc,j_m)$ is the vector of net reaction fluxes and each net reaction flux $j_r$ is the difference between the respective forward and backward reaction fluxes, i.e., $j_r = j_r^+ - j_r^-$.
The choice of a kinetic model amounts to choosing $j_r^+$ and $j_r^-$ as functions of $x \in \RN$.
To each reaction, a chemical force is associated as $f_r := \log j^+_r - \log j^-_r$ \cite{ge2016,rao2018}.

The entropy production rate $\dot \Sigma$ is given by the scalar product between $j$ and $f$ \cite{ge2016,rao2018}, and it is lower bounded by the (twofold) pseudo entropy production rate $\dot \Pi$:
\begin{align}
    \frac{1}{2} \dot \Sigma := \sum_r j_r \log \frac{j^+_r}{j^-_r} \geq  \sum_r \frac{j_r^2}{j^+_r + j^-_r} =: \dot \Pi.
\end{align}
The two mutually inverse diagonal matrices
\begin{align}
    \gJ &:= \textrm{diag}\left[ \frac{1}{j^+_r + j^-_r} \right], \\
    \gF &:= \textrm{diag}\left[ {j^+_r + j^-_r} \right] = \gJ^{-1}
\end{align}
play a prominent role throughout this letter as metric tensors.
Thus, for a positive semidefinite matrix $g$, the $g$-bilinear product is denoted as $\langle v,w \rangle_g := v^T g w$ and the corresponding (squared) norm as $\| v \|_g^2 := \langle v,v \rangle_g$.
As an example, the pseudo entropy production rate is the squared norm of the flux vector with respect to $\gJ$, i.e.,
\begin{align}
    \dot \Pi = j^T \gJ j = \| j \|_{\gJ}.
\end{align}
The net current of the $i$th chemical is given, by \eqnref{eq:dynamics}, as 
\begin{align}
    v_i = \sum_r S_{ir} j_r = (Sj)_i.
\end{align}
As derived in \cite{yoshimura2021TUR}, the diffusion coefficients in the Kramers-Moyal expansion of the chemical master equation are given by
\begin{align} \label{eq:diffusion}
    D_{ij} = \sum_r S_{ir}S_{jr} (j^+_r + j^-_r) = [S \gF S^T]_{ij}.
\end{align}
These quantities govern the fluctuations of the fluxes $v_i$ in the large volume limit.

\paragraph*{Main result.}
The main result of this letter is the following matrix inequality (i.e., the difference of the two matrices is positive semidefinite)
\begin{align} \label{eq:main}
    \gJ \geq S^T D^{\dagger} S,
\end{align}
where $D^{\dagger}$ is the Moore-Penrose inverse of the diffusion matrix $D$.
This inequality is proven in a later section using the Hessian geometry of flux and force spaces.
As a consequence of (\ref{eq:main}), by multiplying with $j^T$ on the left and $j$ on the right, one obtains the scalar inequality
\begin{align} \label{eq:mainTUR}
    \dot \Pi = j^T \gJ j \geq (Sj)^T D^{\dagger} (Sj) = v^TD^{\dagger}v.
\end{align}
The term $v^TD^{\dagger}v$ on right hand side is the generalized inverse of the relative current fluctuations.
To see this, consider the case of a diagonal diffusion matrix.
Then $v^TD^{\dagger}v$ becomes $\sum_{i=1}^n \frac{v_i^2}{D_{ii}}$, which is an expression more commonly encountered in TUR (cf, e.g., \cite{yoshimura2021TUR}).
The inequality (\ref{eq:main}) generalizes this more common expression to non-diagonal and possibly singular diffusion matrices.
Moreover, for any weight vector $d = (d_1,\dotsc,d_m)$, one can form the weighted flux vector $j^d = (j_1d_1,\dotsc,j_md_m)$ and obtain the weighted pseudo entropy production rate $\dot \Pi^d = \|j^d\|_{\gJ}$ as an upper bound for the inverse of the relative fluctuations in $v$ caused by the flux vector $j^d$, given by $(Sj^d)^T D^{\dagger} (Sj^d)$.

One particularly useful choice of weight vectors are the canonical vectors $d = e_r = (0,\dotsc,0,d_r = 1,0,\dotsc,0)$.
In this case, the corresponding flux vectors are denoted as $j^r := (0,\dotsc,0,j_r,0,\dotsc)$ and the pseudo entropy production rate of the $r$th reaction bounds the inverse of relative current fluctuations caused by the $r$th reaction
\begin{align} \label{eq:main2}
    \dot \Pi ^r := \frac{j_r^2}{j_r^+ + j_r^-} \geq (Sj^r)^T D^{\dagger} (Sj^r).
\end{align}
In other words, each reaction $r$ leads to an inverse imprecision in the operation of the CRN given by $(Sj^r)^T D^{\dagger} (Sj^r)$.
Steady state fluxes $j^{ss}$ are characterized by $Sj^{ss} = 0$ and thus the right hand side of the inequality (\ref{eq:mainTUR}) vanishes for all steady states, making it trivially $\dot \Pi \geq 0$.
But even in this case, the multivariate nature of the main inequality (\ref{eq:main}) provides nontrivial results via the inequality (\ref{eq:main2}).
When summing (\ref{eq:main2}) over all reactions, one obtains the following nontrivial bound on the sum of inverse relative current fluctuations caused by all reactions
\begin{align} \label{eq:main3}
    \dot \Pi \geq \sum_r (Sj^r)^T D^{\dagger} (Sj^r).
\end{align}

\begin{figure}
    \includegraphics[width= 0.42\textwidth]{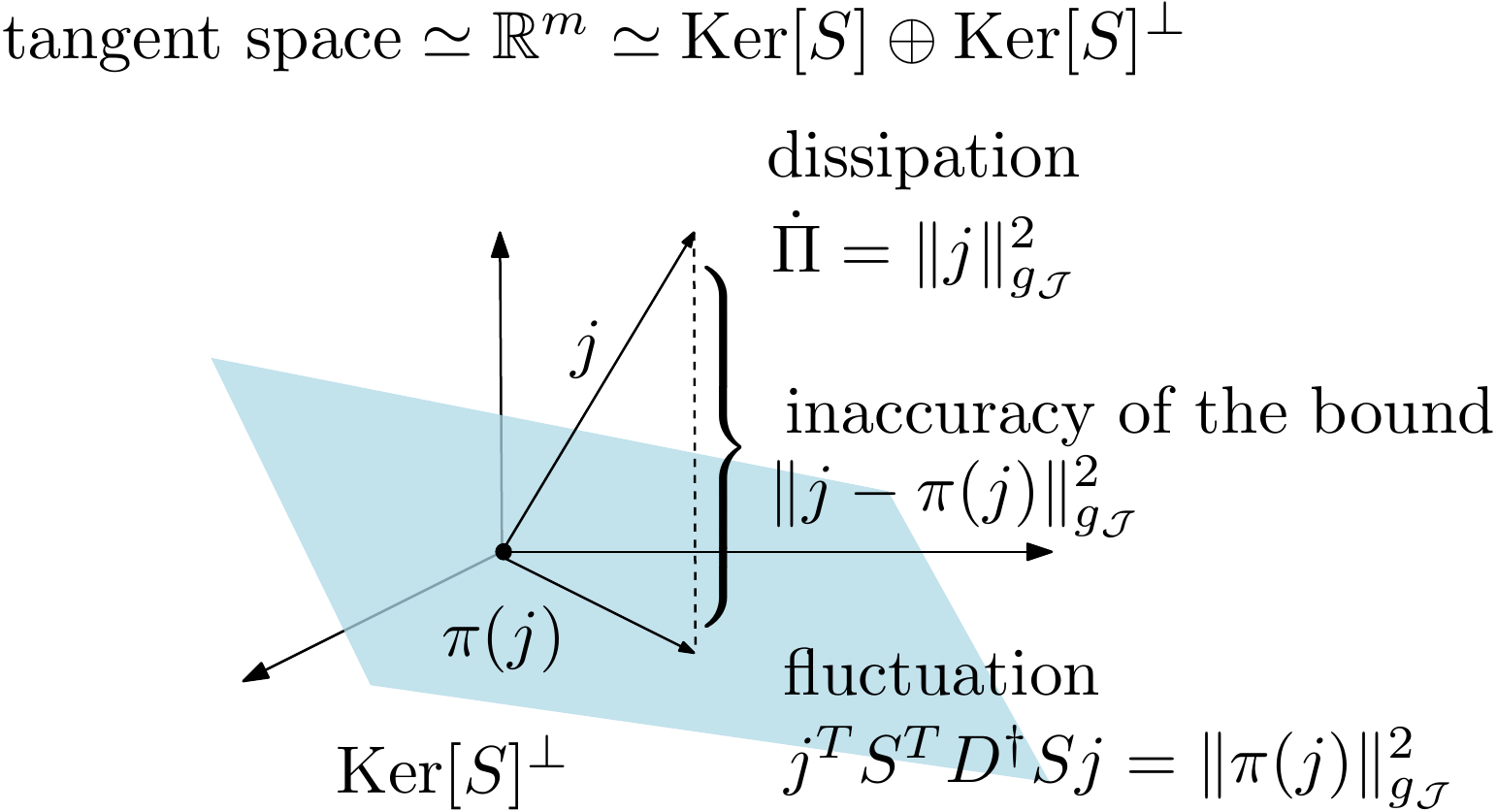}
    \caption{Decomposition of a flux vector $j$ into its $\kerr[S]$-component $j - \pi(j)$ and the respective $\gJ$-orthogonal.
    The squared norms of the three vectors reflect the dissipation, the inverse relative current fluctuations, and the inaccuracy of the bound.
    }
    \label{fig:main_illustration}
\end{figure}

\paragraph*{Geometric interpretation, error estimates, and entropy production decomposition.}
The matrix inequality (\ref{eq:main}) follows from the comparison of two Riemannian metric tensors on the tangent space of the flux space which is isomorphic to $\mathbb{R}^m$.
The space $\mathbb{R}^m$ splits into the linear subspace $\kerr[S]$ and its orthogonal complement $\kerr[S]^{\perp}$ with respect to the metric $\gJ$.
Let $\pi: \RR \rightarrow \kerr[S]^{\perp}$ denote the $\gJ$-orthogonal projection.
Then, for any flux vector $j \in \RR$ \footnote{Note that the algebra derived from the matrix inequality $\gJ \geq S^T D^{\dagger} S$ works for any vector $j \in \RR$, which is taken to be a flux vector.
However, mathematically this should naturally be a tangent vector $\partial j$ to a flux vector.}, the matrix inequality has the following geometrical interpretation:
The pseudo entropy production is given by the squared norm $\dot \Pi = \| j\|^2_{\gJ}$, whereas the inverse relative current fluctuations are quantified by the squared norm $(Sj)^T D^{\dagger} Sj = \|\pi( j) \|_{g_{\mathcal{J}}}^2$ of the $\kerr[S]^{\perp}$-component of $j$.
Thus the error in the estimate of the inverse relative current fluctuations by the pseudo entropy production rate is given by the squared norm of the $\kerr[S]$-component of $j$, i.e., by $\| j - \pi(j) \|_{g_{\mathcal{J}}}^2$.
This is illustrated in Fig. \ref{fig:main_illustration}.
Therefore, depending on the situation, either the $\kerr[S]$-component or the orthogonal component of the individual reaction fluxes $j^r$ should be minimized.
If high precision is required, then $\|\pi( j^r) \|_{g_{\mathcal{J}}}^2$ should be maximized for all reactions $r$ simultaneously.
If, however, exploration is required, then $\|\pi( j^r) \|_{g_{\mathcal{J}}}^2$ should be minimized.
Interestingly, the autocatalytic cores of CRN for growing systems are characterized by nonsingular square stochiometric matrices $S$ and thus posses a trivial kernel space \cite{blokhuis2020}.
This implies that for such CRN, the inequality (\ref{eq:main}) is always an equality and thus that they have the smallest possible relative current fluctuations, i.e., they operate with the highest possible precision.

The splitting of a flux $j$ as $j = \pi(j) + (j - \pi(j))$ gives a pseudo steady state flux $j^{pss} = (j - \pi(j))$ and its $\gJ$-orthogonal component $\pi(j)$, which vanishes at the steady state.
Therefore, one can the spilt the pseudo-entropy production rate into a house keeping $\dot \Pi^{hk}$ and an excess part $\dot \Pi^{ex}$ as
\begin{align}
 \dot \Pi =\| j -\pi(j) \|^2_{\gJ} +  \| \pi(j) \|^2_{\gJ}  =: \dot \Pi^{hk} + \dot \Pi^{ex}.
\end{align}
The excess contribution vanishes at steady states and this splitting is very closely related to the splitting introduced in \cite{kolchinsky2022}.
With respect to TUR, the excess part quantifies the inverse relative current fluctuations.

\paragraph*{Legendre duality between flux and force spaces and Hessian geometry.}
In order to prove the inequality (\ref{eq:main}), we construct a minimal Hessian geometric framework required for the proof.
This geometry (but not its Riemannian aspects) is laid out in more detail in \cite{kobayashi2022graph}.

For any particular architecture of a CRN, it is important to understand its properties under a variety of kinetic and thermodynamic models because fluctuations in the environmental conditions cause variations in the respective models.
The most general setup is to allow, for each $x \in \X$, the flux vector $j \in \J_x \simeq \RR $ and the force vector $f \in \F_x \simeq (\RR)^*$ to vary freely.
This makes the space of fluxes $\J$ and the space of forces $\F$ into vector bundles on the concentration space $\X$.
Both spaces can be formally identified with trivial $\RR$-bundles but they carry more structure as $\J$ lives above the tangent bundle $\T \X$ and $\F$ below the cotangent bundle $\T^* \X$:
\begin{equation} \label{diag:XJF}
\begin{tikzcd}[nodes in empty cells]
    \J \ar[rr,shift left=.55ex,"\textrm{Legendre dual}"] \ar[d, "S"] & & \F \ar[ll,shift left=.55ex,"\textrm{Legendre dual}"] \\
    \T \X \ar[r]  & \X &  \T^*X \ar[u, "S^T"] \ar[l].
\end{tikzcd}
\end{equation}
From macroscopic fluctuation theory it has been derived that there is a Legendre duality between the flux and force bundles induced by the dissipation functions
\begin{align}
        &\Phi(j) = 2\omega^T \left( \textrm{diag}\left[ \frac{j}{2}\right] \textrm{sinh}^{-1} \left( \frac{j}{2} \right) - \left[ \sqrt{1 + \left( \frac{j}{w} \right)} -1 \right] \right)      \\
        \label{eq:Phi_f}
        &\Phi^*(f) = 2 \omega^T \left[ \textrm{cosh} \left( \frac{f}{2} \right) -1 \right],
\end{align}
where $\omega$ is the frenetic activity vector given by the componentwise geometric means between $j_+$ and $j_-$, i.e., $\omega_r = 2\sqrt{j^+_r  j^-_r}$ and the functions and fractions are applied componentwise \cite{mielke2014,mielke2017,renger2018}.
Concretely, the transformation is given by 
\begin{alignat}{2} \label{eq:LD_fullf}
        f &= \nabla_j \Phi(j), &&\quad f_r = \frac{\partial \Phi(j)}{\partial j_r} \\
        \label{eq:LD_fullj}
        j &= \nabla_f \Phi^*(f), &&\quad j_r = \frac{\partial \Phi^*(f)}{\partial f_r}.
\end{alignat}
%= \langle j,S^T \mu \rangle = \langle S j,\mu \rangle$, where $\mu \in \F$ is the chemical potential.  
Concrete physical models enter into Diagram (\ref{diag:XJF}) as follows: 
The kinetics is determined by a section $\X \rightarrow \J, x \mapsto j(x) = j^+(x) - j^-(x)$ and the corresponding force vector is given by the section $\X \rightarrow \F, x \mapsto f(x) = \log j^+(x) - \log j^-(x)$.
The chemical potential vector is given by the section $X \rightarrow \T^* X, x \mapsto \mu(x) = \sum_i \frac{\partial \phi(x)}{\partial x_i} \dd x_i$, where $\phi(x)$ is the free energy function.
A similar differential geometrically flavored viewpoint has been introduced in \cite{oster1974}.

\begin{figure}
    \includegraphics[width= 0.5\textwidth]{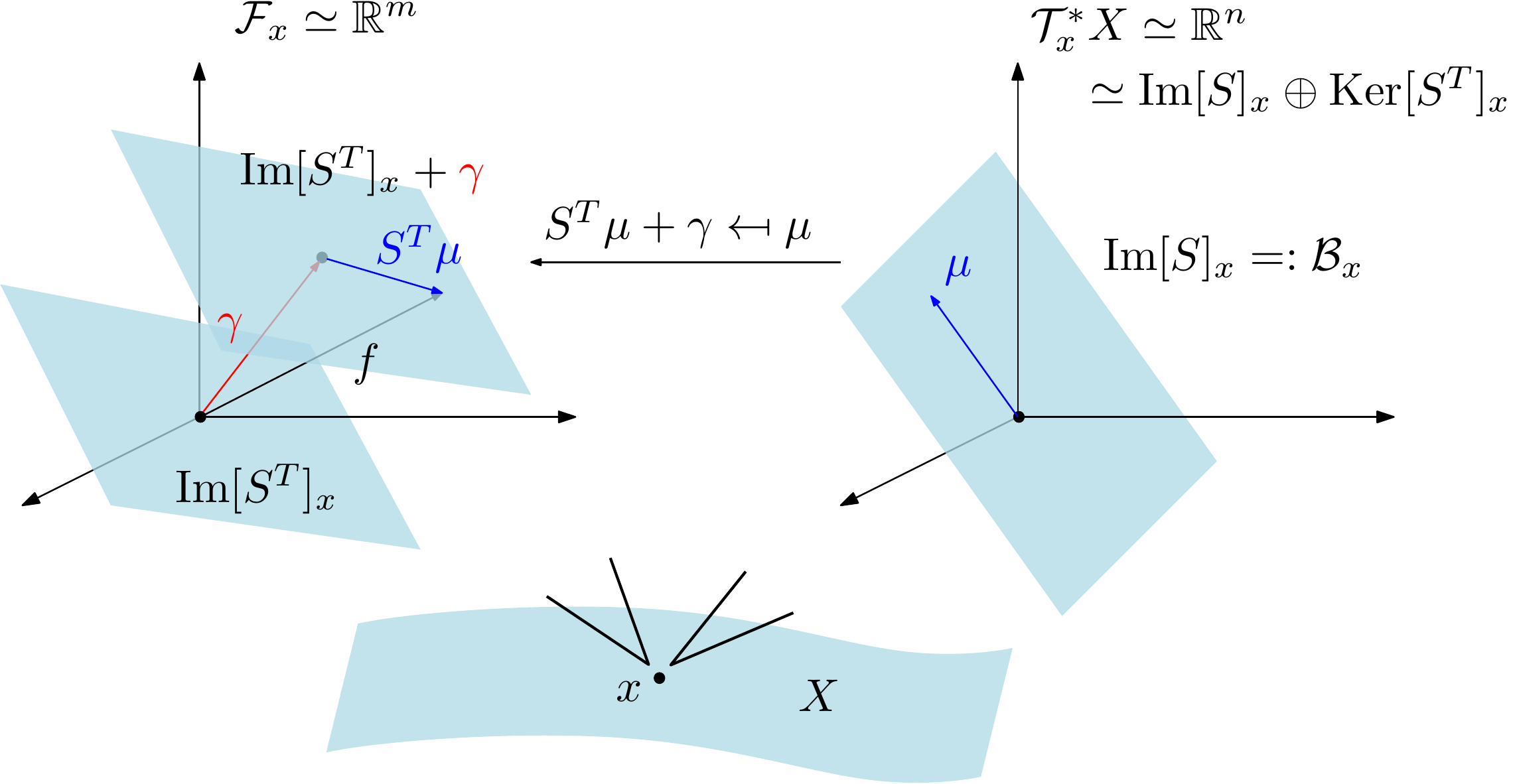}
    \caption{The geometry of the force bundle $\F$.
    The force vector has a contribution $\gamma \in \kerr[S]$, which results from the coupling of the CRN to external reservoirs.
    For mass action kinetics, $\gamma$ is constant over all fibers.
    The chemicals of the CRN determine a potential via $ x \mapsto \mu(x) = \sum_i \frac{\partial \phi(x)}{\partial x_i} \dd x_i$ which can, without loss of generality be assumed to lie in the bundle $\B := \img[S] \subset \T^* \X$.
    The dynamics of the CRN takes place on the affine subbundle $\img[S] + \gamma \subset \F$ as $\mu$ varies depending on $x \in X$.
    The figure shows the respective fibers above some $x \in X$.
    }
    \label{fig:geom}
\end{figure}

The structure of the $\F$-bundle is now analyzed in more detail and its connection to the thermodynamics of the CRN is established.
The bundle $\F$ has an orthogonal decomposition as $\F \simeq \img[S^T] \oplus \kerr[S]$ and a section from $\X$ to $\F$ is given by
\begin{align}
    x \mapsto f = S^T \mu +  \gamma
\end{align}
with $\mu \in \T^* \X$ and $\gamma \in \kerr[S]$.
The base vector $\gamma$ is the vector of cycle affinities which result from the coupling of the CRN to an external reservoir \cite{schnakenberg1976}.
For the dynamics under mass action kinetics this vector remains constant \cite{dal2023geometry}.
The potential $\mu \in \T^* \X$ is the contribution of the chemicals of the CRN to the total chemical force.
Without loss of generality, the potential $\mu$ can be taken inside $\img[S] \subset \T^* \X$ as there is an orthogonal decomposition $\T^* \X \simeq \kerr[S^T] \oplus \img[S]$.
The subbundle of $\T^* \X$ which contains the potential vectors $\mu$ is denoted by $\B$.
Fig. \ref{fig:geom} shows this setup in the fibers $\F_x$ and $\T^*_x \X$ over an arbitrary $x \in \X$.
\iffalse
The CRN admits an equilibrium state for mass action kinetics if and only if the $\kerr[S]$-component $\gamma$ of the section $x \mapsto f = S^T \mu +  \gamma$ vanishes.
%In this case, this section compatible with the section $\X \rightarrow \F$, i.e., $f = S^T \mu$.
The natural bilinear pairing between $\J$ and $\F$ gives the entropy production rate as $\dot \Sigma = \sum_r j_rf_r =: \langle j,f \rangle$.
For a system which admits an equilibrium state, this pairing agrees with the natural pairing between $\T \X$ and $\T^* \X$ as $\dot \Sigma = \langle j,f \rangle = \langle j,S^T \mu \rangle = \langle Sj, \mu \rangle$.
In this case, the steady state condition $Sj = 0$ implies the vanishing of the entropy production, i.e., $\dot \Sigma = \langle Sj, \mu \rangle = 0$.
\fi

\begin{figure}
    \includegraphics[width= 0.5\textwidth]{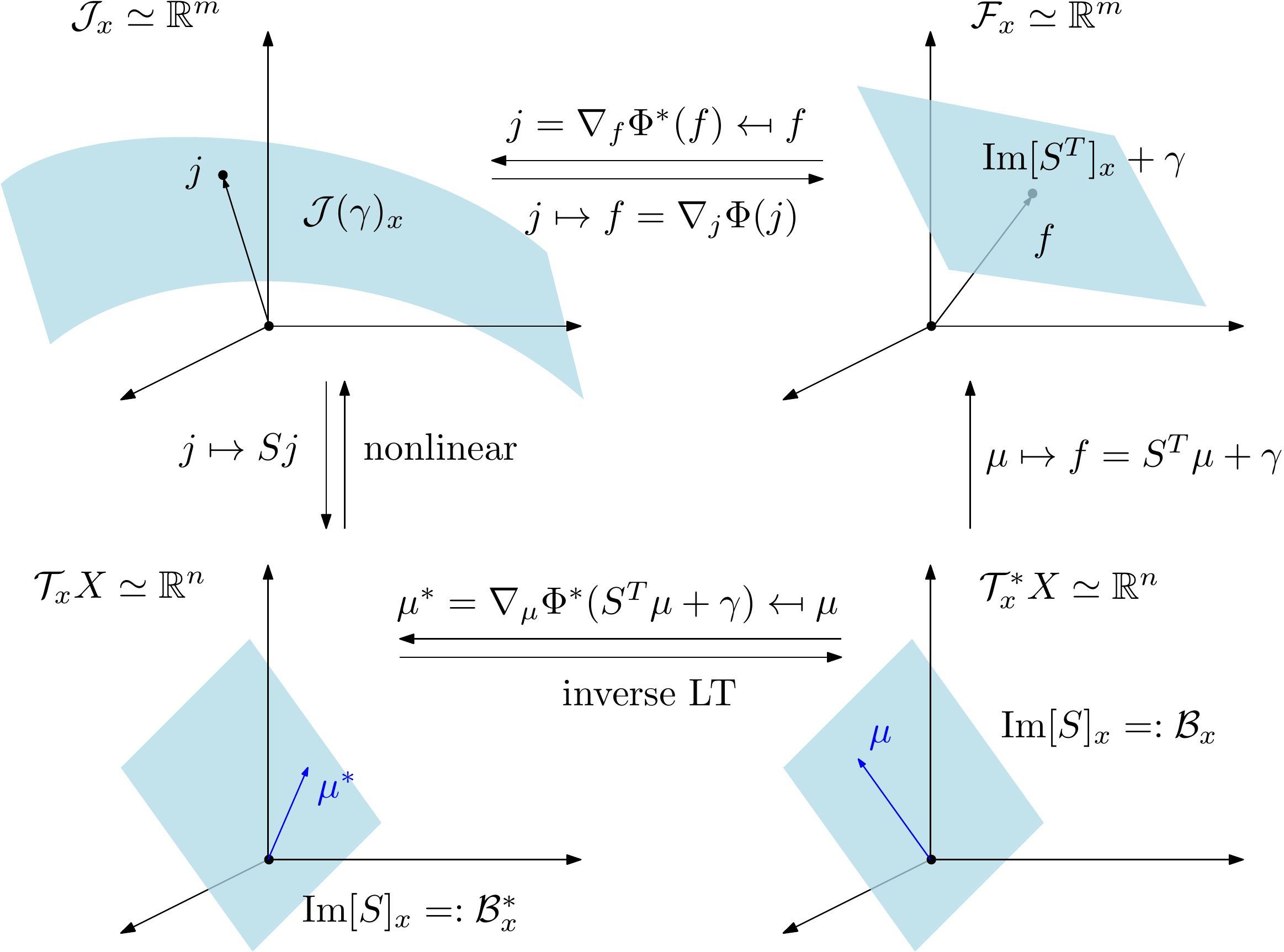}
    \caption{The Hessian geometry of the affine subbundle $\img[S] + \gamma$ of $\F$, its Legendre dual subbundle $\J(\gamma)$ of $\J$ and the respective linear parameter bundles $\B$ and $\B^*$.
    The figure shows the respective fibers above some $x \in X$.
    }
    \label{fig:Hessian}
\end{figure}

From now on, fix a base vector $\gamma \in \kerr[S]$ - this is necessary for the geometric considerations but the final result does not depend on the particular choice of $\gamma$.
Under the Legendre duality (\ref{eq:LD_fullj}), the affine subbundle $\img[S^T] + \gamma \subset \F$ transforms into a nonlinear subbundle $\J(\gamma)$ of $\J$.
The package of Hessian geometry provides a (nonlinear) parametrization of this subbundle where the Legendre transform $\B^*$ of $\B$ acts as the parameter space (see \cite{kobayashi2022geometry,kobayashi2022graph} or the Supplementary Information for more details).
The coordinates $\mu^* \in \B^*_x$ are obtained by using the dissipation function $\Phi^*(S^T \mu + \gamma)$, which was introduced in \eqnref{eq:Phi_f} :
\begin{align}
    \mu^* := \nabla_{\mu} \Phi^*(S^T \mu + \gamma), \quad \mu^*_i = \frac{\partial \Phi^*(S^T \mu + \gamma)}{\partial \mu_i}.
\end{align}
The fibers $\B^*_x$ of the bundle $\B^*$ are isomorphic to $\img[S]_x \subset \RR$ \footnote{More precisely, $\B^*_x$ is isomorphic to $(\img[S])^* \subset (\RR)^*$ but this for the sake of readability, we not distinguish between finite dimensional vector spaces and their duals because they are (noncanonically) isomorphic.} and the diffeomorphism $\B^* \rightarrow \J(\gamma)$ is nonlinear and has no closed expression in general.
However, its inverse is given by the linear map
\begin{align}
    S: J(\gamma) \rightarrow \B, \quad j \mapsto \mu^* = Sj.
\end{align}
(Again, details can be found in \cite{kobayashi2022geometry,kobayashi2022graph} or the Supplementary Information).
This leads to the commutative diagram of diffeomorphisms and diffeomorphic embeddings shown in Fig. \ref{fig:Hessian}.

\paragraph*{Metric Structure.}
The dissipation functions $\Phi(j)$ and $\Phi^*(f)$ equip the fibers $\J_x$ and $\F_x$ with Riemannian metrics via the respective Hessian matrices
\begin{align}
    (g_{\J})_{rs} = g_{\J}\left(\frac{\partial}{\partial j_r},\frac{\partial}{\partial j_s}\right) = \frac{\partial ^2 \Phi(j)}{\partial j_r \partial j_s},
\end{align}
for which the shorthand notation $g_{\J} = \frac{\partial ^2 \Phi(j)}{\partial j \partial j}$ will be used.
The proofs of the facts stated here follow directly from the chain rule but are given in the Supplementary information for the sake of completeness.
The matrix representing the Hessian metric on $\F_x$ is inverse to $g_{\J}$
\begin{align}
    g_{\F} := \frac{\partial ^2 \Phi^*(f)}{\partial f \partial f} = g_{\J}^{-1},
\end{align}
and the Legendre duality between $\F$ and $\J$ is an isometry.
Moreover, the linear parametrization $\B_x \rightarrow \img[S^T]_x + \gamma \subset \F_x$ is an isometric embedding with the Riemannian metric on $\B_x$ given by
\begin{align}
g_{\B} = \frac{\partial^2 \Phi(S^T \mu + \gamma)}{\partial \mu \partial \mu} = [Sg_{\F}S^T] = D,
\end{align}
where for the last equality the expression \eqnref{eq:diffusion} for the matrix of diffusion coefficients was used.
Using the fact that the Hessian matrices representing Riemmannian metrics of Legendre dual coordinates are mutually Moore-Penrose inverses of each other yields the Riemannian metric on $\B^*$ as
\begin{align}
g_{\B^*} = g_{\B}^{\dagger} = D^{\dagger}.
\end{align}
The pullback of $g_{\B^*}$ to $\J(\gamma)_x$ is given by
\begin{align}
g_{\J(\gamma)} = S^Tg_{\B}^{\dagger}S = S^TD^{\dagger}S.
\end{align}
Due to the isometry of the diagram in Fig. \ref{fig:Hessian}, this means that the metric $S^TD^{\dagger}S$ coincides with $\gJ$ on $J(\gamma)_x$ and vanishes on its $\gJ$-orthogonal complement.
This proves the matrix inequality (\ref{eq:main}).

\paragraph*{Discussion.}
In this letter, a multivariate TUR for CRN is derived based on Hessian geometry.
This novel geometrical understanding allows to understand the inverse relative current fluctuations as the square norm of the $\kerr[S]^{\perp}$-component of the flux vector.
The error in estimating the magnitude of the inverse fluctuations by the pseudo entropy production rate is therefore given by the square norm of the $\kerr[S]$-component of the flux vector and this geometry therefore dictates how to minimize or maximize the current fluctuations in CRN by tuning the direction of the respective reaction vectors.

Moreover, the multivariate nature of the TUR allows to quantify the relative current fluctuation even in the steady state.

More importantly, the constancy of the base vector $\gamma$ implies that the CRN dynamics takes place in an affine subbundle of the force bundle and, by Legendre duality, in a (nonaffine) subbundle of the flux bundle, therefore restricting the fluctuations of the fluxes.
This is the reason for the discrepancy between dissipation and inverse relative current fluctuations.
It will be interesting to see how this transfers to other nonequilibrium systems.

Finally, the geometrical understanding of TUR sets the stage for going beyond the Fokker-Planck approximation to the chemical master equation.
This, however, requires Hessian geometry on the infinite dimensional spaces of force and flux trajectories.

\begin{acknowledgments}
We thank Atsushi Kamimura and Shuhei Horiguchi for valuable discussions.
This research is supported by JSPS KAKENHI Grant Numbers 19H05799 and 21K21308, and by JST CREST JPMJCR2011 and JPMJCR1927.
\end{acknowledgments}

\iffalse

\appendix

\section{Pullback} \label{app:geodesics}

Pullback and pushforward might be not so common in physics but they correspond to covariance and contravariace as follows:...

\fi

\bibliography{apssamp}% Produces the bibliography via BibTeX.

\end{document}